\documentclass[a4paper]{article}
\usepackage{amsmath,amsfonts,epsfig}

\newcommand{\Rl}{\mathbb{R}}

\newcommand{\nb}{\bar{n}}

\newcommand{\dt}{\! \cdot \!}
\newcommand{\wdg}{\! \wedge \!}

\newcommand{\deriv}[2]{\frac{\partial #1}{\partial #2}}
\newcommand{\alp}{\alpha}

\newcommand{\lam}{\lambda}

\newcommand{\nn}{\nonumber}
\newcommand{\half}{{\textstyle \frac{1}{2}}}
\newcommand{\Rrev}{\tilde{R}}

\newcommand{\la}{\langle}
\newcommand{\ra}{\rangle}
\newcommand{\clg}{{\mathcal{G}}}

\begin{document}

\noindent
{\bf\Large \textsf{Circle and sphere blending with conformal \\
geometric algebra}} 

\vspace{0.2cm}

\noindent
{\large Chris Doran\footnote{E-mail: C.Doran@mrao.cam.ac.uk}}

\vspace{0.2cm}

\noindent
Astrophysics Group, Cavendish Laboratory, Madingley Road, \\
Cambridge CB3 0HE, UK.

\vspace{0.4cm}

\begin{abstract} Blending schemes based on circles provide smooth
`fair' interpolations between series of points. Here we demonstrate a
simple, robust set of algorithms for performing circle blends for a
range of cases.  An arbitrary level of $G$-continuity can be achieved
by simple alterations to the underlying parameterisation.  Our method
exploits the computational framework provided by conformal geometric
algebra. This employs a five-dimensional representation of points in
space, in contrast to the four-dimensional representation typically
used in projective geometry. The advantage of the conformal scheme is
that straight lines and circles are treated in a single, unified
framework.  As a further illustration of the power of the conformal
framework, the basic idea is extended to the case of sphere blending
to interpolate over a surface.

\vspace{0.4cm}

\end{abstract}

\section{Introduction}

In a range of applications we often seek curves and surfaces that have
an aesthetically pleasing `roundedness' to them.  One way to make this
concept concrete is through looking for globally-optimised `minimum
variation curves'~\cite{mor92}.  The philosophy behind this idea is
straightforward.  We usually prefer curves that are close to circular
over curves with sharp turns.  This is particularly true when
designing camera trajectories, where sudden changes in curvature can
have a very disorienting effect.  Circular paths are characterised by
having constant curvature, so a natural idea in forming interpolations
between control points is to find a curve that minimises the total
change in curvature.  The problem with such a strategy is that these
curves can be extremely hard to compute.  If one adopted a variational
strategy, with endpoint conditions, the equations for the curve can be
as high as fifth order and are even more difficult to treat than those
of elasticity.  Such equations can only be solved numerically and do
not have straightforward, controllable, analytic solutions.  The
problem is even more acute if multiple control points are involved,
as even numerical computation can be extremely difficult.

A more straightforward, local scheme that provides smooth
interpolations was introduced by Wenz~\cite{wenz96} and later extended
by Szilv\'{a}si-Nagy \& Vendel~\cite{szil00}.  The idea explored by
these authors is to generate curves that are as close as possible to
circles.  Given four points $X_0, \ldots, X_3$ we construct the circles
$C_1$ through $X_0$, $X_1$ and $X_2$, and $C_2$ through $X_1$, $X_2$
and $X_3$.  The curve between $X_1$ and $X_2$ is then formed by
smoothly interpolating between the two circles.  This idea was further
extended by S\'{e}quin \& Yen~\cite{seq01} and S\'{e}quin \&
Lee~\cite{seq03}, who introduced an angle-based circle blending
scheme.  The angle-based scheme gives better results than the earlier,
midpoint scheme, and we argue here that it is geometrically the
`correct' one.  S\'{e}quin \& Lee also showed how to achieve
$G_2$-continuity (and higher order continuity, if desired), and
demonstrated the value of angle-based blending for interpolating over
the surface of a sphere.

In this paper we further explore the geometry associated with circle
blending, following the methods developed by S\'{e}quin and his
coworkers.  The essential idea is that the natural way to transform
between two circles is via a \textit{conformal} transformation.
Conformal transformations leave angles invariant, but can alter
distances.  Euclidean transformations are the subset of conformal
transformations that also leave distances invariant.  Conformal
transformations in a plane can take any three chosen points to any
three image points.  As such, they can transform a line or circle into
any other circle.  In this geometry, straight lines are examples of
circles that pass through the point at infinity.  By exploiting the
features of conformal geometry, we can write robust code that treats
(straight) lines and circles in a single, unified manner.  This
eliminates the need to check for special cases.  Similarly, in three
dimensions, planes and spheres are treated as examples of the same
object. So a single routine can interpolate between points on a
sphere, and will reduce to the planar case when four points happen to
lie in a plane.

To fully exploit the advantages of conformal geometry we work in the
mathematical framework of \textit{geometric
algebra}~\cite{gap,DLL-sheff}.  This algebra treats points, lines,
circles, planes and spheres, and the transformations acting on them,
in a unified algebraic framework.  A number of authors have argued for
the advantages of the \textit{conformal geometric algebra} framework
for computer graphics
applications~\cite{DLL-sheff,LL-surf,agacse-proc,SIG2000,SIG2001}.
The present work should be viewed in this context. We show how complex
problems such as finding the conformal transformation between a line
and a circle reduce to simple, robust expressions in the geometric
algebra framework.  As a further application we show how the same
framework naturally extends to sphere-blending over a surface.  This
suggests a new method of characterising surfaces that does not
require the concept of swept curves.

This paper starts with an introduction to conformal geometric algebra.
This introduction is self-contained, but to keep its length down a
number of concepts are introduced with a minimum of explanation.  We
then turn to the question of how to mathematically encode
transformations between circles.  We find the conformal transformation
that achieves this and explore its properties.  Some subtleties
involving the orientation of the transformation are explained, and we
demonstrate how they are easily resolved in the conformal framework.
We then provide a series of examples of blended curves, and illustrate
the effects of demanding higher-order $G$-continuity.  We finish by
introducing a method of sphere blending and discuss the potential of
this idea for encoding surfaces.

\section{Conformal geometry and Euclidean space}

The starting point for our description of geometry is the conformal
group.  This marks a radical departure from conventional descriptions
of Euclidean space based on projective geometry and homogeneous
coordinates.  The main advantage of basing the description in a
conformal setting is that \textit{distance} is encoded simply, making
the geometry well suited to describing the real three-dimensional
world.  

Suppose we start with an $n$-dimensional Euclidean vector space
$\Rl^n$.  The conformal group consists of the set of all
transformations of $\Rl^n$ that leave angles invariant.  These include
translations and rotations, so the conformal group includes the set of
Euclidean transformations as a subgroup.  The conformal group on
$\Rl^n$ has a natural representation in terms of rotations in a space
two dimensions higher, with signature $(n+1,1)$.  So, in the same way
that projective transformations are linearised by working in a space
one dimension higher than the Euclidean base space, conformal
transformations are linearised in a space two dimensions higher.  The
conformal representation of points in Euclidean three-space consists
of vectors in a 5-dimensional space.  While this may appear to be an
unnecessary abstraction, working in this five-dimensional space does
bring a number of advantages.

To exploit the conformal representation we need a standard
representation for a Euclidean point in the five-dimensional conformal
space.  Given that we are occupying a space two dimensions higher, two
constraints are required to specify a unique point.  The first of
these is that our underlying representation is homogeneous, so $X$ and
$\lam X$ represent the same point in Euclidean space.  The second
constraint is that the vector $X$ is \textit{null},
\begin{equation}
X^2 =0.
\end{equation}
(The existence of null vectors is guaranteed by the fact that the
conformal vector space has mixed signature.)  This is essentially the
only further constraint that can be enforced which is consistent with
homogeneity and invariant under orthogonal transformations in
conformal space.  Now suppose that $e_1$, $e_2$ and $e_3$ represent
three vectors in the three-dimensional base space, and we add to these
the vectors $e_0$ and $e_4$.  These satisfy
\begin{equation}
e_0^2=-1, \qquad e_4^2 = +1,
\end{equation}
and all 5 vectors $\{e_0, \ldots, e_4\}$ are orthogonal.  From the two
extra vectors we define the two null vectors $n$ and $\nb$ by
\begin{equation}
n = e_4+e_0, \qquad \nb = e_4 - e_0, \qquad n \dt \nb = 2.
\end{equation}
(It is a straightforward exercise to confirm that these two vectors
both have zero magnitude.)  From these we need to chose a vector to
represent the origin.  This is conventionally taken as $-\half \nb$.
The vector $X$ can therefore be written as
\begin{equation} 
X = 2x - \nb + \alpha n
\end{equation}
where $x$ is the equivalent three-vector representation of the point
in Euclidean space.  The variable $\alpha$ must be chosen so that $X$
is null, which fixes $\alpha = x^2$.  We therefore arrive at the
following representation of a point in conformal space:
\begin{equation}
X = 2x + x^2 n - \nb.
\label{cfv}
\end{equation}
This representation can be arrived at using more geometrical reasoning
by considering a stereographic projection~\cite{gap,DLL-sheff}.  It is
immediately clear from equation~\eqref{cfv} that $n$ represents the
point at infinity.

\begin{figure}
\begin{center}
\includegraphics[width=5cm]{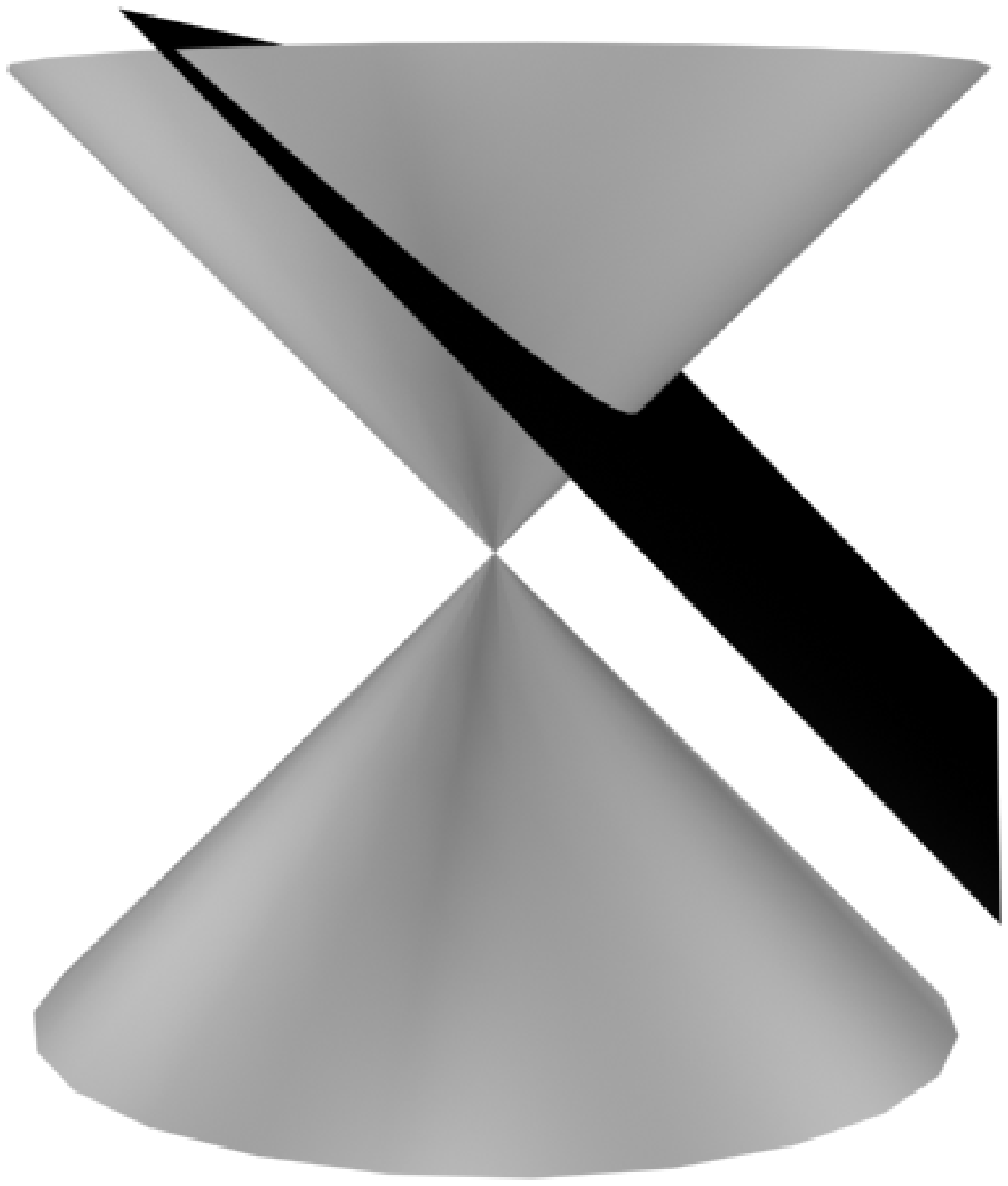}
\includegraphics[width=4cm]{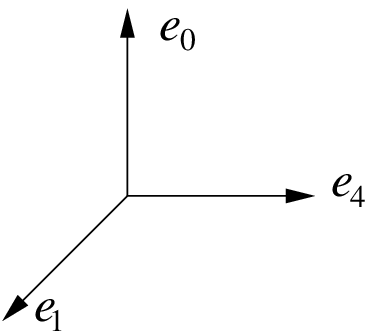}
\end{center}
\caption[The null cone]{\textit{The null cone.}
In a three dimensional space of signature $(2,1)$ the set of null
vectors form a cone.  To ensure a unique representation of points
that maintains orientation, we restrict to the subsection with
positive $e_0$ component.  The set of vectors satisfying $X \cdot n = -2$
defines the `standard' representation of points, in this case defining
the conformal representation of a one-dimensional space.}
\label{Fnullcone}
\end{figure}

The Euclidean coordinates of the point $x$ are recovered from $X$ via
the homogeneous relation
\begin{equation}
x_i = - \frac{X \dt e_i}{X \dt n}, \quad i = 1, 2, 3.
\label{xifromX}
\end{equation}
This relation confirms that $X$ and $\lambda X$ define the same
Euclidean point.  In principle, one could allow for $\lam$ to be
negative, but in practice this should be avoided.  Restricting to
positive $\lam$ is is equivalent to stating that $X$ must always have
a positive $e_0$ component.  The reason for this restriction can be
seen in figure~\ref{Fnullcone}.  The equation $X^2=0$ generates a cone
structure, and we only want to employ one half of this to represent
points.  Maintaining a positive $e_0$ component throughout all
algorithms enables us to keep track of orientations consistently.

Given a point $a$, represented by the unnormalised vector $A$, we can
place $A$ in the standard form of equation~\eqref{cfv} with the map
\begin{equation}
A \mapsto -2 \frac{A}{A \dt n}.
\end{equation}
One should be wary of employing this map when writing code, as the
right-hand side is singular if $A$ happens to be the point at
infinity.  As is the case for projective geometry, it is better
practice to let the normalisation run free, and only use
equation~\eqref{xifromX} in the final stage to recover the
coordinates.

The power of the conformal representation starts to become clear when
we consider the inner product of points.  Suppose that $X$ and $Y$ are
the conformal representations of the points $x$ and $y$ respectively,
both in the standard form of equation~\eqref{cfv}.  Their inner
product is
\begin{align}
X \dt Y &= \bigl( x^2n + 2x -\nb  \bigr) \dt \bigl( y^2n + 2y -
\nb \bigr) \nn \\
&=-2x^2 -2y^2 +4x\dt y \nn \\
&= -2(x-y)^2.  
\label{confinner}
\end{align}
This is the essential result that underpins the conformal approach to
Euclidean geometry.  The inner product in conformal space encodes the
\textit{distance} between points in Euclidean space.  This is the
reason why points are represented with null vectors --- the distance
between a point and itself is zero.

Generalising equation~\eqref{confinner} to unnormalised vectors, the
distance between points can be written
\begin{equation}
|x-y|^2 = -2 \frac{X \dt Y}{X \dt n \, Y \dt n}.
\label{Eucdist}
\end{equation}
Any transformation of $X$ and $Y$ that leaves this product invariant
must represent a symmetry of Euclidean space.  Because such a
transformation must leave the inner product invariant, it must be an
orthogonal transformation in conformal space.  Such a transformation
ensures that null vectors remain null vectors, and so continue to
represent points.  The transformation must also leave $n$ invariant if
the distance is to remain unchanged.  We can now see that the
Euclidean group is the subgroup of conformal transformations that
leaves invariant the point at infinity.  In what follows we will find
applications for both Euclidean and more general conformal
transformations.

\section{Conformal Geometric Algebra}

\textit{Geometric algebra} was introduced by the nineteenth century
mathematician W.K.~Clifford, and has found many applications in
physics~\cite{gap,hes-nf1}.  Often in this work the name
\textit{Clifford algebra} is used, but for applications in geometry it
is becoming increasingly common to see Clifford's original name of
geometric algebra to describe the field.  A geometric algebra is
constructed over a vector space with a given inner product. The
\textit{geometric} product of two vectors $a$ and $b$ is defined to be
associative and distributive over addition, with the additional rule
that the square of any vector is a scalar,
\begin{equation}
aa = a^2 \in \Rl.
\end{equation}
If we write
\begin{equation}
ab + ba = (a+b)^2 - a^2 - b^2
\end{equation}
we see that the symmetric part of the geometric product of any two
vectors is also a scalar.  This defines the inner product, and we
write
\begin{equation}
a \dt b = \half(ab+ba).
\end{equation}
The remaining, antisymmetric part of the geometric product returns the
\textit{outer} or \textit{exterior} product familiar from projective
geometry.  We write this as
\begin{equation}
a \wdg b = \half(ab-ba).
\end{equation}
The totally antisymmetrised sum of geometric products of vectors
defines the exterior product in the algebra.

The geometric product of two vectors can now be written
\begin{equation}
ab = a \dt b + a \wdg b.
\end{equation}
Under the geometric product, orthogonal vectors anticommute and
parallel vectors commute.  The product therefore encodes the basic
geometric relationships between vectors.  Now that we know how to
multiply vectors together, it is straightforward to construct the
entire geometric algebra of a given vector space.  This is facilitated
by introducing an orthonormal frame of vectors $\{e_i\}$.  These
satisfy
\begin{equation}
e_i \dt e_j = \eta_{ij},
\end{equation}
where $\eta_{ij}$ is the metric tensor.  For a space of signature
$(p,q)$, $\eta_{ij}$ is a diagonal matrix consisting of $p$ $+1$s and
$q$ $-1$s.  The space of interest to us here is the conformal space of
signature $(4,1)$.  A basis for this is provided by the vectors $e_0,
\ldots, e_4$, with $e_0$ having negative square.  The algebra
generated by these vectors has 32 terms in total, and is spanned by
\begin{equation*} 
\begin{array}{rcccccc}
& 1 & \{e_i \} & \{e_i \wdg e_j \}  &  \{ e_i \wdg e_j \wdg e_k \} & \{I
e_i \} & I \\ 
\mbox{grade} & 0 & 1 & 2 & 3 & 4 & 5\\
\mbox{dimension} & 1  & 5  & 10 & 10  & 5 & 1.
\end{array}
\end{equation*} 
We refer to this algebra as $\clg(4,1)$.  The term `grade' is used to
refer to the number of vectors in any exterior product.  The
dimensions of each graded subspace are given by the binomial
coefficients.

The highest grade term in $\clg(4,1)$ is called the
\textit{pseudoscalar} and is given the symbol $I$.  This is defined by
\begin{equation}
I = e_0 e_1 e_2 e_3 e_4 .
\label{pseud}
\end{equation}
The pseudoscalar commutes with all elements in the algebra, a feature
of odd-dimensional algebras, and the $(4,1)$ signature of the space
implies that the pseudoscalar satisfies
\begin{equation}
I^2 = -1.
\end{equation}
So, algebraically, $I$ has the properties of a unit imaginary.  But it
also plays a definite geometric role, as multiplication by the
pseudoscalar performs a duality transformation in conformal space.  A
matrix representation of $\clg(4,1)$ can be constructed in terms of
$4\times 4$ complex matrices.  These can be found in the physics
literature in the guise of the Dirac matrices~\cite{itz-quant}.  For
practical applications this matrix representation has little value and
one is better off coding up the algebraic rules explicitly.

A general element of $\clg(4,1)$ is called a \textit{multivector} and
can consist of a sum of terms all grades in the algebra.  Arbitrary
elements of $\clg(4,1)$ can be added and multiplied together.  A
multivector that consists only of terms of a single grade is said to
be \textit{homogeneous}.  The geometric product of a pair of
homogeneous multivectors decomposes as follows:
\begin{equation}
A_r B_s = \la A_r B_s \ra_{|r-s|} + \la A_r B_s \ra_{|r-s|+2} + \cdots
+ \la A_r B_s \ra_{r+s}.
\end{equation}
The angle brackets $\la M \ra_r$ are used to denote the projection
onto the grade-$r$ terms in $M$.  The dot and wedge symbols are used
to generalise the inner and outer products to the lowest and highest
grade terms in a general product:
\begin{align}
A_r \dt B_s &= \la A_r B_s \ra_{|r-s|} \nn \\ 
A_r \wdg B_s &= \la A_r B_s \ra_{r+s} ,
\end{align}
where $A_r$ and $B_s$ are homogeneous multivectors of grade $r$ and
$s$ respectively.  One only ever uses the dot and wedge symbols when
applied to homogeneous multivectors.  Of particular importance are the
inner and outer products with a vector.  These satisfy
\begin{align}
a \dt A_r &= \half(a A_r - (-1)^r A_r a) \nn \\
a \wdg A_r &= \half(a A_r + (-1)^r A_r a).
\end{align}
Some straightforward algebra confirms that
the outer product is associative~\cite{gap}.

An important algebraic operation applied to multivectors is
\textit{reversion}.  This plays an analogous role to transposition in
matrix algebra.  The reverse of a multivector $M$ is denoted by
$\tilde{M}$ and is defined by reversing the order of all geometric
products of vectors in $M$.  An arbitrary multivector in $\clg(4,1)$
can be written as
\begin{equation}
M = \alpha + a + B + T + Ib + I \beta
\end{equation}
where $\alpha$ and $\beta$ are scalars, $a$ and $b$ are vectors, $B$
is a bivector and $T$ is a trivector.  The reverse of $M$ is then
given by
\begin{equation}
\tilde{M} = \alpha + a - B - T + Ib + I \beta.
\end{equation}

A conformal transformation in the Euclidean space $\Rl^3$ is
represented by an orthogonal transformation in $\clg(4,1)$.  Of
particular relevance here are special conformal transformations.
These have determinant $+1$, and correspond to orientation-preserving
transformations on $\Rl^3$.  In geometric algebra a special conformal
transformation, applied to an arbitrary multivector $A$, can be
written in the compact form
\begin{equation}
A \mapsto R A \Rrev.
\end{equation}
Here $R$ is an even-grade element (grades 0, 2 and 4) satisfying 
\begin{equation}
R \Rrev = 1.
\end{equation}
Because our representation is homogeneous, this normalisation
constraint can usually be relaxed to allow for $R\Rrev$ to be an
arbitrary positive scalar.  The element $R$ is called a
\textit{rotor}.  As an example, the (unnormalised) rotor for the
Euclidean translation between $x$ and $y$ can be written
\begin{equation}
T_{xy} = (n\dt Y) nX + (n \dt X) Yn,
\end{equation}
where $X$ and $Y$ are the conformal equivalents of $x$ and $y$
respectively.  Every rotor can be written in term of the exponential
of a bivector as
\begin{equation}
R = \pm \exp(-B/2).
\end{equation}
The bivector $B$ is the generator of the rotation, and is an element
of the Lie algebra of the conformal group.

\section{Geometric primitives in conformal algebra}

We have now established how points in $\Rl^3$ are represented by null
vectors in the conformal algebra $\clg(4,1)$.  We now need to establish
how higher dimensional objects are represented.  Given a null vector
$Y$, one view of the point this represents is as the solution space of
the equation
\begin{equation}
Y \wdg X = 0, \qquad X^2 = 0.
\end{equation}
where here $X$ is treated as a vector variable.  This is a similar
idea to projective geometry, and generalises immediately to
higher-grade objects.  Given a multivector $A_r$, the outer product of
$r$ distinct vectors, the geometric object that this represents is the
solution space of
\begin{equation}
A_r \wdg X = 0, \qquad X^2 =0.
\label{solspc}
\end{equation}
This pair of equations is clearly homogeneous in $X$.  If a conformal
transformation is applied to $A_r$, taking it to $RA_r\Rrev$, then the
solution space transforms in exactly the same way (that is, $X \mapsto
R X \Rrev$).  In this manner conformal transformations are easily
applied to higher-grade objects.

After vectors, the next objects to consider are bivectors.  Given a
bivector $B$ (formed from the outer product of two vectors), the
solution space of equation~\eqref{solspc} depends on the sign of
$B^2$.  If $B^2$ is negative there are no solutions, if $B^2=0$ there
is one solution, and if $B^2$ is positive there are two solutions.
This final case corresponds to the situation where $B = X \wdg Y$,
where $X$ and $Y$ are a pair of null vectors.  Given a bivector of
this form it is straightforward to recover $X$ and $Y$~\cite{gap}, so
a bivector can encode a pair of points.

Next consider a trivector $L$.  The main example of relevance is the
trivector formed from the outer product of three points $X_1$, $X_2$
and $X_3$, so
\begin{equation}
L = X_1 \wdg X_2 \wdg X_3.
\end{equation}
Any vector $X$ that is a solution of $L \wdg X = 0$ must be a linear
combination of $X_1$, $X_2$ and $X_3$.  The fact that $X$ is null and
has arbitrary scale implies that in $\Rl^3$ the solution space is
one-dimensional.  To see what this space is, consider the null vector
$C$, where
\begin{equation} 
C = L n L.
\end{equation}
If $X$ satisfies $X\wdg L = 0$, then $X$ commutes with $L$.  It
follows that 
\begin{equation}
\frac{X \dt C}{ X \dt n \, C \dt n} = \frac{L^2}{C \dt n} =
\mbox{constant}.
\end{equation}
The solution set $X$ therefore consists of points at an equal distance
from some point $C$.  It follows that the trivector $L$ represents the
\textit{circle} through $X_1$, $X_2$ and $X_3$.  If this circle also
passes through the point at infinity, the line is straight.  That is,
the test that a line $L$ is straight is that
\begin{equation}
L \wdg n = 0.
\end{equation}
Similarly, the straight line through $X_1$ and $X_2$ is given by
\begin{equation}
L = X_1 \wdg X_2 \wdg n.
\end{equation}
Conformal algebra therefore treats lines and circles in a unified
manner as trivectors in $\clg(4,1)$.  This is to be expected, as a
conformal transformation can always transform a line into a circle.

The radius $\rho$ of the circle $L$ is found from
\begin{equation}
\rho^2 = - \frac{L^2}{(L \wdg n)^2}.
\end{equation}
So again we see how a metric object (the radius) is recovered from a
homogeneous representation of a geometric entity (the circle $L$).
The angle $\theta$ between two circles (or lines) is found from
\begin{equation}
\cos\theta = \frac{L_1 \dt L_2}{|L1| \, |L2|}.
\end{equation}
As expected, this is invariant under the full conformal group.

Similar considerations apply to the 4-vector $S$ defined by the four
points $X_1, \ldots ,X_4$,
\begin{equation}
S = X_1 \wdg X_2 \wdg X_3 \wdg X_4.
\end{equation}
The object defined by $S$ is the unique sphere through the four points
(which cannot all lie on a line or circle if $S\neq 0$).  The centre
of the sphere is given by $SnS$, and if the centre lies at infinity
the sphere reduces to a plane.  As with the case of a straight line,
the plane defined by the three points $X_1$, $X_2$ and $X_3$ is
\begin{equation}
P = X_1  \wdg X_2 \wdg X_3 \wdg n.
\end{equation}

Lines, circles, planes and spheres can all be intersected in a
straightforward manner in conformal geometric algebra.  The two cases
of greatest significance are those of a circle and a sphere, and of
two spheres.  Treating the latter first, any two spheres intersect in
a circle (provided they touch).  This reduces to a line if both
spheres are in fact planar.  The intersection $L$ is found simply by
\begin{equation}
L = (I S_1) \dt S_2,
\end{equation}
which is a trivector, as required.  Similarly, the intersection of a
circle $C$ and a sphere $S$ is given by
\begin{equation}
B = (IC) \dt S.
\end{equation}
This is a bivector, as a circle can intersect a sphere or plane in
zero, one, or two points.  In the case where $C$ is a straight line and
$S$ is a flat plane, one of the points of intersection is at infinity.
As we proceed we will see how the orientations implicit in lines and
surfaces are neatly encoded in these intersection formulae.

\section{Circle blending}
\label{Scblnd}

\begin{figure}
\begin{center}
\includegraphics[angle=-90,width=8cm]{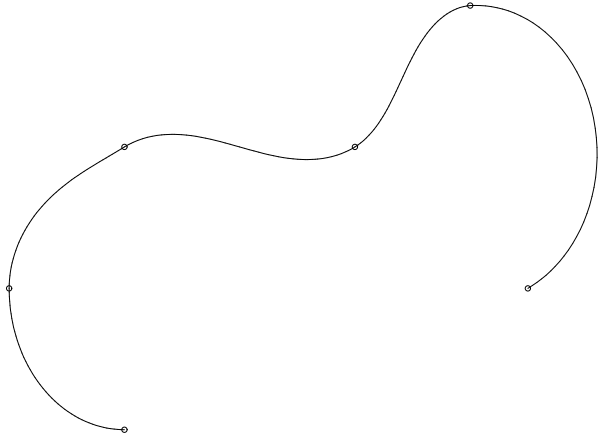}
\end{center}
\caption[Circles splines in the plane]{\textit{Circles splines in the
plane.}  The curve through each intermediate point is generated by
blending between the two circles at each end, producing a smooth curve
with limited change in its curvature from point to point.  The curve
shown has $G_2$ continuity, using the scheme described in
section~\ref{Scont}. }
\label{Fcircbl1}
\end{figure}

Suppose we are given a series of $n+1$ points $x_0, \ldots, x_n$ and we
seek a smooth curve through all of these points.  The idea behind the
circle blending scheme is to find a curve that is as close as possible
to a circle and which passes through all of the points.  A typical
setup is illustrated in figure~\ref{Fcircbl1}.  Consider the four
points $x_{i-1} \ldots x_{i+2}$.  We form the circle $C_i$ through
$x_{i-1}$, $x_{i}$ and $x_{i+1}$, and the circle $C_{i+1}$ through
$x_{i}$, $x_{i+1}$ and $x_{i+2}$.  In the region between $x_i$ and
$x_{i+1}$ we need a curve that blends smoothly between the circles
$C_i$ and $C_{i+1}$.

\begin{figure}
\begin{center}
\includegraphics[width=5cm]{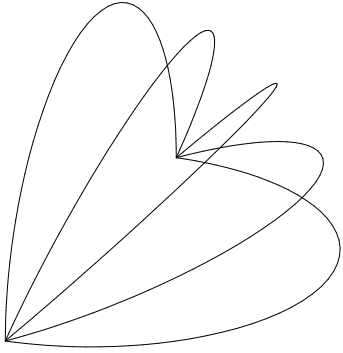}
\end{center}
\caption[Interpolating between a pair of
circles]{\textit{Interpolating between a pair of circles.}  The two
outer circles share two points in common, and between them specify a
sphere.  All of the intermediate circles also lie on this sphere.}
\label{F2crcs}
\end{figure}

A range of schemes exist for interpolating between circles, but one is
naturally picked out from the viewpoint of conformal geometry.  We are
presented with two circles $C_1$ and $C_2$, both of which share two
points in common:
\begin{align}
C_1 &= X_0 \wdg X_1 \wdg X_2 \nn \\
C_2 &= X_1 \wdg X_2 \wdg X_3.
\end{align}
Given the (conformal) points $X_1$, $X_2$, and the circles $C_1$ and
$C_2$, there are no further objects to specify.  First we need to
define  family of circles between $C_1$ and $C_2$.  This is easily
achieved by finding the transformation between them.  The
transformation must map between two circles, so is a conformal
transformation.  The only bivector generator for such a transformation
is the antisymmetrised product between $C_1$ and $C_2$.  A
transformation governed by this generator is simply a rotation in
$\clg(4,1)$.  If we normalise the circles by defining
\begin{equation}
\hat{C_i} = \frac{C_i}{|C_i|}
\end{equation}
then the interpolated circle is simply
\begin{equation}
\hat{C}_{12}(\lam) = \frac{1}{\sin(\theta)} \left(
\sin\bigl((1-\lam)\theta\bigr) \hat{C}_1 + \sin(\lam\theta) \hat{C}_2
\right). 
\label{crcblnd}
\end{equation}
Here $\theta$ is the angle between the circles, defined by
\begin{equation}
\cos(\theta) =  \hat{C}_1 \dt \hat{C}_2.
\end{equation}
(The case of $\theta > \pi$ will be addressed shortly.)  This method
of interpolating between circles recovers the angle-blending scheme
of S\'{e}quin \& Yen~\cite{seq01}, thus providing a firm geometric
reason for preferring that scheme.  At this point there is no reason
for $C_1$ and $C_2$ to lie in the same plane.  If they lie on
different planes then the four points $X_0, \ldots, X_3$ define a
sphere.  Since the interpolated circles can be written
\begin{equation}
\hat{C}_{12}(\lam) = \frac{1}{\sin(\theta)} X_1 \wdg X_2 \wdg \left(
\sin\bigl((1-\lam)\theta\bigr) \frac{X_0}{|C_1|}
+ \sin(\lam\theta) \frac{X_2}{|C_2|} \right),
\end{equation}
we see that all points on these circles are combinations of $X_0
\ldots X_3$.  It follows that all of the intermediate circles also lie
of the sphere defined by $X_0, \ldots , X_3$, which is a desirable
property for a range of applications~\cite{seq03}.  The basic
interpolation scheme is illustrated in figure~\ref{F2crcs}

\begin{figure}
\begin{center}
\includegraphics[angle=-90,width=6cm]{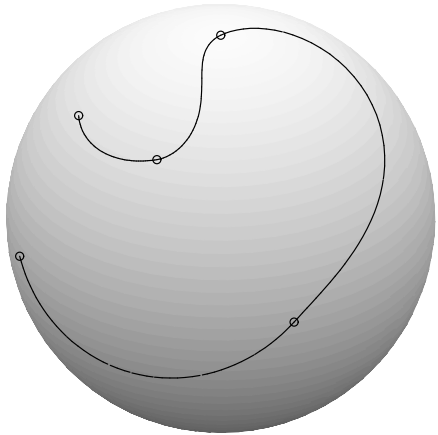}
\end{center}
\caption[Circles splines on a sphere]{\textit{Circles splines on a sphere.}
The curve is generated using the same algorithm as for the planar
case.  The algorithm ensures that all intermediate points in the
middle section lie on the sphere specified by the four control
points.  In the example, all five control points lie on the same
sphere, so the curve sits smoothly on the sphere.}
\label{Fsphplt}
\end{figure}

Now that we have a straightforward means of encoding the circle
blends, we need to parameterise the actual trajectory between the
points.  A simple means for achieving this is to start with the
straight line between $X_1$ and $X_2$.  This path is parameterised by
\begin{equation}
Y_{12}(\lam) = -(1-\lam) X_2 \dt n \, X_1 - \lam X_1 \dt n \, X_2 +
\lam (1-\lam) X_1 \dt X_2 \, n,
\end{equation}
and the line itself is described by
\begin{equation}
L_{12} = X_1 \wdg X_2 \wdg n.
\end{equation}
The rotor that transforms between the line $L_{12}$ and the circle
$C_{12}(\lam)$ is simply
\begin{equation}
R_{12}(\lam) = 1+ \hat{C}_{12}(\lam) \hat{L}_{12}.
\label{Rtblnd}
\end{equation}
So once the (normalised) blended circle $\hat{C}_{12}(\lam)$ is found,
the path itself is given by transforming from the straight line to the
circle.  The path between $X_1$ and $X_2$ is therefore simply
\begin{equation}
X_{12}(\lam) = R_{12}(\lam) Y_{12}(\lam)  \Rrev_{12}(\lam).
\label{ptblnd}
\end{equation}
This is extremely simple to code up.  We never have to calculate the
centre of a circle, and the algorithm deals equally easily with
straight lines or circles.  The algorithm keeps all intermediate
points on the sphere defined by each group of four points.  If all of
the base points lie on the same sphere, then we compute smooth curves
over this sphere.  An example of this is shown in
figure~\ref{Fsphplt}.

\section{Problems and special cases}

Provided the sequence of any four points are unique, the circles to be
blended and the intermediate straight line are all well defined.  The
only aspects of the algorithm that can be problematic are concerned
with the definition of the angle $\theta$ in equation~\eqref{crcblnd}.
There are two cases over which care must be taken, both of which can
be treated fairly easily.  The first obvious problem is that the
definition of the blended segment breaks down if $\sin(\theta)=0$, which
occurs when $\cos(\theta)=+1$ and when $\cos(\theta)=-1$.  The former
case corresponds to $\theta=0$ and implies that the two curves are the
same.  This is easily caught by setting a threshold value for $\theta$
below which eqn~\eqref{crcblnd} is replaced by the small-$\theta$
limit of
\begin{equation}
C_{12}(\lam) = (1-\lam) \hat{C}_1 + \lam \hat{C}_2.
\end{equation}

\begin{figure}
\begin{center}
\includegraphics[angle=-90,width=6cm]{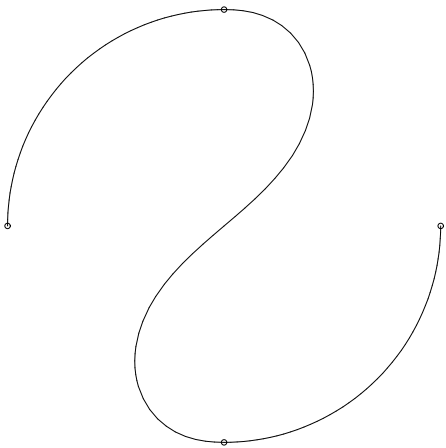}
\end{center}
\caption[Four points on a circle]{\textit{Four points on a circle}.
The four points lie on a circle, but the two circles we interpolate
between are equal and opposite.  An interpolation scheme in the plane
is quite possible, but the case is pathological.  If one point moves
slightly off the plane, the interpolated curve will jump to a
spherical case.}
\label{Fpathol}
\end{figure}

The case of $\theta=\pi$ is more complicated, and corresponds to a
somewhat pathological example.  This occurs when the two circles only
differ in their orientation, so $ \hat{C}_2=- \hat{C}_1$.  A sample
configuration is shown in figure~\ref{Fpathol}.   All four points lie
in a plane, so the interpolated curve must also lie in this plane.  It
is possible to define an interpolation scheme for this case.  If we
let $P$ denote the plane containing the circles, then the generator of
the transformation must act in the $P$ plane, leaving $X_1$ and $X_2$
invariant.  The bivector generator for the transformation is therefore
\begin{equation}
B_{12} = (X_1 \wdg X_2) P.
\end{equation}
This produces the type of interpolated curve shown in
figure~\ref{Fpathol}.  The problem here is that this case is not
smoothly connected to the general case.  Imagine moving one of the
control points in figure~\ref{Fpathol} slightly out of the page.  The
four points then define a sphere, and the interpolated curve will lie
on this sphere.  This is a totally different curve to the planar case.
If we are only interested in planar plots, then everything is
well-behaved and the above case can be easily dealt with.  But if we
are interested in curves in three-dimensional space, this case is best
avoided altogether with the addition of further control points.

\begin{figure}
\begin{center}
\includegraphics[width=8cm]{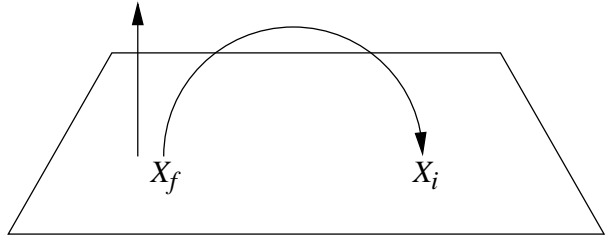}
\end{center}
\caption[Intersection of a plane and a circle]{\textit{Intersection of
a plane and a circle}.  Both the plane and the circle are oriented
objects, and the intersection bivector defined by
equation~\eqref{CPiint} returns $X_f \wdg X_i$.} 
\label{Fintplncirc}
\end{figure}

The case of oppositely oriented circles relates to the second key
issue, which is what happens if the circles are greater than
$180^\circ$ apart?  The arccos function will not return the correct
angle, and the interpolation scheme will effectively blend the wrong
way.  This problem has an elegant solution within the conformal
framework.  The question reduces to finding the correct midpoint
circle $C_{1,2}$.  This is given by
\begin{equation}
C_{1,2} = 
\begin{cases}
\hat{C}_1 + \hat{C}_2, \quad & \theta< \pi \\
-(\hat{C}_1 + \hat{C}_2), \quad & \theta > \pi 
\end{cases}.
\end{equation}
The case of $\theta=\pi$ corresponds to equal and opposite circles,
which is the one awkward case we have to avoid.  All we need is a test
to decide which is the correct mid-circle for the geometry, and the
arccos function can be used to determine $\theta/2$ unambiguously from
\begin{equation}
\cos(\theta/2) = \hat{C}_1 \dt \hat{C}_{1,2}.
\end{equation}
To find a suitable test we consider the plane of points equidistant
from $X_1$ and $X_2$.  This plane is defined by
\begin{equation}
\Pi = I (X_1 \wdg X_2) \dt n.
\end{equation}
All planes are oriented, so the sign here is important.  This choice
corresponds to the normal to the plane (defined by the right-hand
rule) pointing in the direction from $X_1$ to $X_2$.  Now suppose that
we intersect a circle $C$ with the plane $\Pi$.  The result is two
points encoded in the bivector $B$, where
\begin{equation}	
B = I (C \Pi + \Pi C ).
\label{CPiint}
\end{equation}
The bivector $B$ contains the points in the order 
\begin{equation}
B = X_f \wdg X_i,
\end{equation}
where $X_f$ is the point where the circle intersects the plane from
the negative to positive side, and $X_i$ is the opposite point (see
figure~\ref{Fintplncirc}).

Now suppose that $M_1$ and $M_2$ are the midpoints of the circle
segments $C_1$ and $C_2$ respectively.  Both lie in the plane $\Pi$.
The correct mid-circle should have its $X_f$ intersection point closer
to the midpoint of $M_1$ and $M_2$ than its $X_i$ point.  If this is
not the case, then the circle has the wrong orientation.  We therefore
only need form the quantity
\begin{equation}
\alp = \bigl( (M_1 \dt n \, M_2 + M_2 \dt n \, M_1 ) \wdg n \bigr) \dt
B
\end{equation}
where $B$ is given by equation~\eqref{CPiint} with $C$ defined by
\begin{equation}
C = \hat{C}_1 + \hat{C}_2.
\end{equation}
The test is now
\begin{equation}
C_{1,2} = 
\begin{cases}
\hat{C}_1 + \hat{C}_2, \quad & \alp < 0 \\
-(\hat{C}_1 + \hat{C}_2), \quad & \alp > 0
\end{cases}.
\end{equation}
This conditional statement covers all special cases (where various
circles are degenerate or reduce to straight lines) and the border
cases of $\alp=0$ is the single degenerate case mentioned above.
With the mid-circle found, the angle $\theta$ is computed, and the
interpolation scheme of equation~\eqref{crcblnd} can be followed.

\begin{figure}
\begin{center}
\includegraphics[width=6cm]{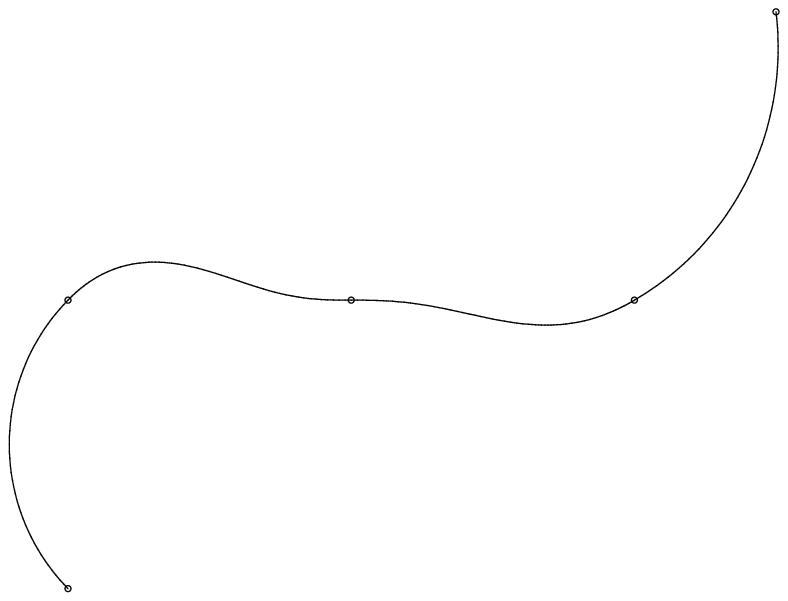}
\includegraphics[width=6cm]{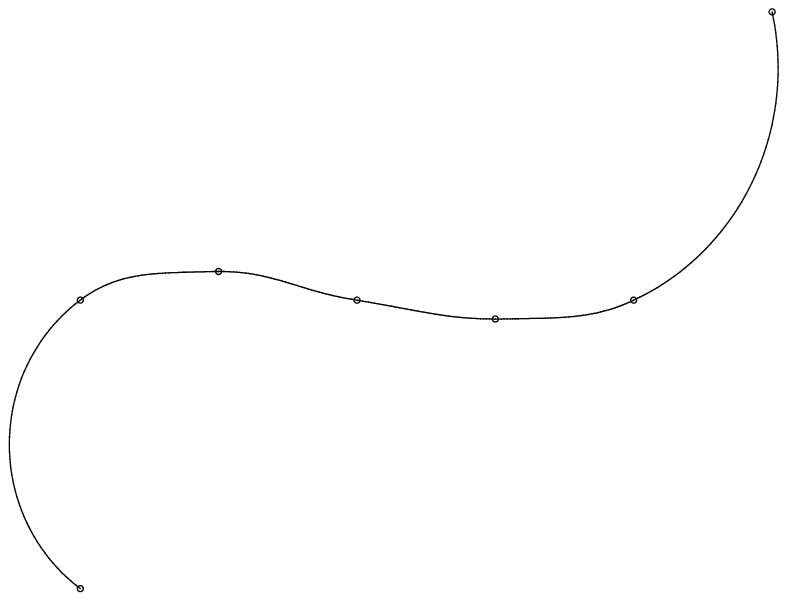}
\end{center}
\caption[An unwanted inversion]{\textit{An unwanted inversion}. The
blending scheme does not always minimise curvature, and can lead to
additional inversions in certain cases, as shown on the left-hand
side. This is because the gradient at any point is determined entirely
by the two points directly next to it.  The problem can be resolved by
introducing additional midpoints in the problem area and re-blending,
producing the smooth curve shown on the right.}
\label{Fprobcase}
\end{figure}

A further problem with the blending scheme is highlighted in
figure~\ref{Fprobcase}.  The geometry of the control points ensures
that an unwanted inversion arises in the middle of the plot.  This is
because the gradient at each control point is fixed entirely by the
two points on either side (see section~\ref{Scont}), and so in this
case is forced to be flat, as illustrated.  The problem can be removed
in a number of ways.  One of the simplest is to define a series of
midpoints, and then repeat the blending algorithm.  The result of a
single iteration of this type is also shown in figure~\ref{Fprobcase}.
Of course, one can go further and continue to recursively define
midpoints, which will also generate a smooth blend.  This latter
approach has the attractive feature that it removes the need for any
trigonometric evaluations, as the mid-circle is found easily using the
method described above.

\section{Continuity}
\label{Scont}

An important issue in any interpolation scheme is the order of
continuity at the control points.  In this section we will only
consider $G$-continuity.  To achieve $C$-continuity some
reparameterisation will have to occur on each spline.  For the case of
a camera fly-by, this can be achieved by specifying the desired
velocity along the curve.

Our blending scheme is defined by equation~\eqref{ptblnd}, and before
we can differentiate this we need a result for the derivative of a
rotor.  The definition of a (normalised) rotor is sufficient to prove
that we can always write
\begin{equation}
\deriv{R}{\lam} = - \half R B
\end{equation}
where $B$ is a bivector, which in general will also be a function of
$\lam$.  (The factor of $-1/2$ is a useful convention).  To find the
tangent vector to the (conformal) curve $X(\lam)$ we form the line
\begin{equation}
T = X \wdg X' \wdg n
\end{equation}
where the dash denotes the derivative with respect to $\lam$.  To see
how this behaves we return to equation~\eqref{ptblnd} and form the
derivative at $\lam=0$. For convenience we will assume that the rotor
in equation~\eqref{Rtblnd} has been normalised, though this is
unimportant when forming the tangent vector.  In evaluating the
derivatives we can use the results that
\begin{equation}
R(\lam) X_1 \Rrev(\lam) = X_1, \qquad
R(\lam) X_2 \Rrev(\lam) = X_2.
\end{equation}
Differentiating these, we see that
\begin{equation}
X_1 \dt B(\lam) = X_2 \dt B(\lam) = 0,
\end{equation}
and these also hold for derivatives of $B$ with respect to $\lam$.

On differentiating~\eqref{Rtblnd} we find that (dropping the
subscripts on $R$ and $Y$)
\begin{equation}
X' = R (Y \dt B + X_2 \dt n \, X_1  - X_1 \dt n \, X_2 +(1-2 \lam) X_1
\dt X_2 \, n) \Rrev.
\end{equation}
It follows that
\begin{equation} 
\left. X' \right|_{\lam=0} =  X_2 \dt n \, X_1  - X_1 \dt n \,
X_2 +   X_1 \dt X_2 \,  R_0 n \Rrev_0,
\end{equation} 
where $R_0=R(\lam=0)$.  The tangent vector at $X_1$ is therefore given
by (up to a scale factor)
\begin{align}
T_1 &=   X_1 \dt n \, X_2 \wdg X_1 \wdg n -  X_1 \dt
X_2 \,  (R_0 n \Rrev_0) \wdg X_1 \wdg n  \nn \\
&= - \Bigl( R_0\bigl(( X_1 \wdg X_2 \wdg n) \dt X_1 \bigr) \Rrev_0
\Bigr) \wdg n \nn \\
&= -(C_1 \dt X_1) \wdg n.
\end{align}
But this is precisely the tangent vector to the circle $C_1$ at that
point $X_1$.  So the tangent at each of the control points is defined
by the circle through the control point and its two adjoining points.
The tangent vectors at a connection are therefore continuous, which
ensures $G_1$ continuity.

\begin{figure}
\begin{center}
\includegraphics[width=3cm]{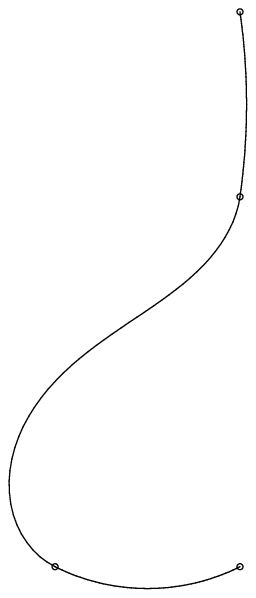}
\hspace{0.5cm}
\includegraphics[width=3cm]{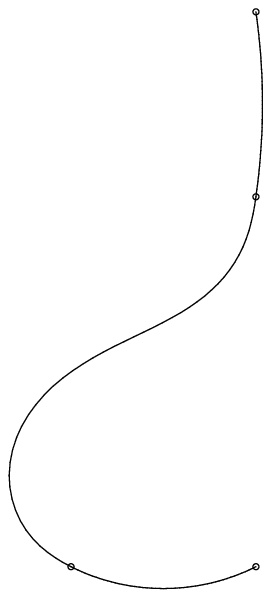}
\end{center}
\caption[Comparison between $G_1$ and $G_2$
  continuity]{\textit{Comparison between $G_1$ and $G_2$
continuity}. The $G_1$ blend is shown on the left, and the $G_2$ blend
on the right.  The $G_2$ blend hugs the control points more closely
than the $G_1$ case, and for most cases the $G_2$ curves are the more
aesthetically satisfying.}
\label{Fg2comp}
\end{figure}

Next we need to consider $G_2$ continuity.  For this we need the
circle
\begin{align}
C_v &= X \wdg X' \wdg X'' \nn \\
&= R \Bigl( Y \wdg  \bigl(Y' + Y \dt B ) \wdg \bigl(Y'' + 2 Y' \dt B + Y \dt
B'  ) \Bigr) \Rrev. 
\end{align}
Evaluating this at $\lam=0$ we obtain
\begin{align}
C_v &= R_0 \, (- X_2 \dt n \, X_1) \wdg ( X_2 \dt n \, X_1 - X_1 \dt n \,
X_2 + X_1 \dt X_2 \, n) \wedge \nn \\ 
& \quad (-2 X_1 \dt X_2 \, n + 2  X_1 \dt X_2 \, n \dt B(0) ) \,
\Rrev_0 \nn \\ 
&=  -2 (X_1 \dt n \,  X_2 \dt n \, X_1 \dt X_2) C_1 \nn \\
& \quad - 2(X_2 \dt n \,
X_1 \dt X_2) \, R_0 \, X_1 \wdg ( - X_1 \dt n \,
X_2 + X_1 \dt X_2 \, n) \wdg ( n \dt B(0)) \,  \Rrev_0.
\end{align}
This contains a term in $C_1$, and an additional term controlled by
the bivector $B$ (evaluated at $\lam=0$).  The first term is the
desired one, as it ensures that the radius of curvature at the control
point is defined by the circle through it.  The second term is not
wanted, and implies that the simple scheme defined by
equation~\eqref{crcblnd} does not guarantee $G_2$ continuity.  This
was first pointed out by S\'{e}quin \& Lee~\cite{seq03}.  To provide
$G_2$ continuity we need only ensure that the derivative of the
interpolating rotor $R$ vanishes at the endpoints.  This is simply
achieved by replacing equation~\eqref{crcblnd} with
\begin{equation}
\hat{C}_{12}(\lam) = \frac{1}{\sin(\theta)} \left(
\sin\bigl((1-3\lam^2+2\lam^3)\theta\bigr) \hat{C}_1 +
\sin((3\lam^2-2\lam^3)\theta) \hat{C}_2 
\right). 
\label{g2bldn}
\end{equation}
This blending scheme does now have $G_2$ continuity, and is the scheme
use to produce the figures in section~\ref{Scblnd}.  In
figure~\ref{Fg2comp} we show a comparison between the two schemes.  As
expected, the $G_2$ scheme smoothes out some of the curvature around
the control points.  One can extend this idea to obtain any desired
order of continuity.  This is achieved by replacing the polynomial in
equation~\eqref{g2bldn} by a higher-order blend.

\section{Sphere blending}

One can straightforwardly apply the ideas developed here to swept
surfaces using a version of the scheme described by Szilv\'{a}si-Nagy
\& Vendel~\cite{szil00}.  But in this section we aim to explore an
alternative idea, based on sphere blending.  A sphere is described as
a 4-vector in the conformal geometric algebra, and these can be
transformed and interpolated in a similar manner to circles.

As a simple example, suppose that the points  $X_1$, $X_2$ and $X_3$
define the vertices of a triangle.  To each corner we attach a sphere,
$S_1$, $S_2$ and $S_3$.  Each sphere passes through the three vertices
of the triangle, and a fourth point which can be viewed as a control
point.  That is, we can write
\begin{align}
S_1 &= A_1 \wdg X_1 \wdg X_2 \wdg X_3 \nn \\
S_2 &= A_2 \wdg X_1 \wdg X_2 \wdg X_3 \nn \\
S_3 &= A_3 \wdg X_1 \wdg X_2 \wdg X_3 .
\end{align}
These assume that none of $A_1$, $A_2$ and $A_3$ lie on the circle
defined by $X_1$, $X_2$ and $X_3$.  We next need to define a blend
over the surface.  For this we introduce the barycentric coordinates
$(\lam, \mu, \nu )$, subject to
\begin{equation}
0 \leq \lam, \mu, \nu \leq 1, \quad \lam + \mu + \nu = 1.
\end{equation}
We let $Y(\lam,\mu)$ denote the conformal representation of the point
in the triangle corresponding to the barycentric coordinates $(\lam,
\mu, 1-\lam-\mu)$.  Similarly, we define a (linear) sphere blend by
\begin{equation}
S(\lam, \mu) = \lam \hat{S}_1 + \mu \hat{S}_2 + \nu \hat{S}_3.
\end{equation}
We could employ a trigonometric blending scheme over the (abstract)
sphere defined by the unit 4-vectors $\hat{S}_1$, $\hat{S}_2$ and
$\hat{S}_1$, but this raises a number of complications.  There is no
single straightforward generalisation of barycentric coordinates over
a spherical triangle, and each alternative scheme has its own merits
and drawbacks~\cite{praun03}.  Here we have adopted the simplest,
linear blending scheme, which generates interesting surfaces.

\begin{figure}
\begin{center}
\includegraphics[width=5cm]{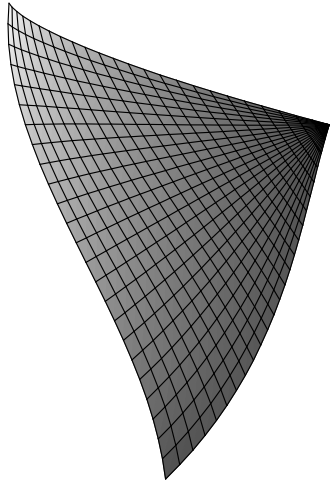}
\end{center}
\caption[A blended surface]{\textit{A blended surface.}
A sphere is attached to each of the vertices of a triangle.  The
surface is defined by a linear blend of the spheres.   }
\label{Fsph}
\end{figure}

Now that we have the sphere and the point on the triangle defined, all
that remains is to define the conformal transformation from the point
to the blended surface.  First we define the plane $P$ through the
three base points,
\begin{equation}
P = X_1 \wdg X_2 \wdg X_3 \wdg n.
\end{equation}
Next we define the rotor $R$ for the conformal transformation between
the plane and the blended sphere,
\begin{equation}
R(\lam,\mu) = 1-\hat{S}(\lam,\mu) \hat{P}.
\end{equation}
Finally, the surface itself is defined by the points $X(\lam,\mu)$,
where
\begin{equation}
X(\lam,\mu) = R(\lam,\mu) Y(\lam,\mu)\Rrev(\lam,\mu).
\end{equation}
A typical surface defined in this manner is shown in
figure~\ref{Fsph}.  The result is aesthetically quite pleasing,
yielding a smooth blend free of sudden changes in curvature.

Much work remains to extend this idea to a complete framework for
defining blends over surfaces.  The control points need to be chosen
so as to reflect the geometry around the triangle, and continuity
between blended spheres may be hard to achieve.  Here we hope to have
demonstrated that such an approach may be feasible.

\section{Summary}

Smooth splines between control points can be defined in terms of
blended circles.  The splines pass through all points, and have no
extra control points.  The natural geometric framework for handling
circles is conformal geometry.  This geometry is encapsulated in a
simple, unified manner by employing the geometric algebra of a
five-dimensional space.  All transformations are easily defined, and
algorithms can be written in such a way as to minimise problems with
special cases.  Similar ideas can be applied to sphere blends over a
surface, and in future work we will explore the potential of this idea
for design and for encoding surface data.

\section*{Acknowledgements}

CD thanks the EPSRC for their support.

\end{document}